\pdfoutput=1
\documentclass[final,authoryear,3p,times]{elsarticle}

\let\citep\cite
\let\citet\cite

\usepackage{etex}
\usepackage{graphicx}
\usepackage{pstricks}
\usepackage{tikz}
\usepackage{subfig}
\graphicspath{{./figures/}}
\usepackage{amsmath,amssymb,amsfonts,amsthm,textcomp}
\usepackage{type1ec,type1cm,fix-cm,lmodern}
\usepackage[utf8]{inputenc}
\usepackage{color}
\usepackage{moreverb}
\usepackage{booktabs}
\usepackage{rotating}
\usepackage{tabularx}
\usepackage[section]{placeins}
\usepackage{xcolor}
\usepackage{stmaryrd}
\usepackage{relsize}
\usepackage{harmony}
\usepackage[normalem]{ulem}

\usepackage{tensor}
\usepackage{bm}
\usepackage{xstring}
\newcommand{\T}[2]{
  \ifcat\noexpand{#1}\relax % check if the argument is a control sequence
    \IfEqCase{#2}{
      {1}{\underline{\bm{#1}}}
      {2}{\underline{\underline{\bm{#1}}}}
      {3}{\underline{\underline{\underline{\bm{#1}}}}}
      {4}{\underline{\underline{\underline{\underline{\bm{#1}}}}}}
    }
 \else
    \IfEqCase{#2}{
      {1}{\underline{\bm{\mathbf{#1}}}}
      {2}{\underline{\underline{\bm{\mathbf{#1}}}}}
      {3}{\underline{\underline{\underline{\bm{\mathbf{#1}}}}}}
      {4}{\underline{\underline{\underline{\underline{\bm{\mathbf{#1}}}}}}}
    }
  \fi
  }

\newcommand{\gradx}[1]{\textbf{grad}\,{#1}}

\newcommand{\divx}[1]{\textbf{div}\,{#1}}

\newcommand{\lapx}[1]{\textbf{lap}\,{#1}}

\newcommand{\tr}[1]{\text{tr}\,{#1}}
\newcommand{\diffo}[1]{\text{d}{#1}}

\newcommand{\partn}[3]{\frac{\partial^{#3}{#1}}{\partial{#2}^{#3}}}
\newcommand{\spart}[3]{\partial^{#3}_{#2}{#1}}

\makeatletter
\def\threevdots{\,\,\vbox{\baselineskip2\p@ \lineskiplimit\z@
  \kern6\p@\hbox{.}\hbox{.}\hbox{.}}\,\,}
\makeatother

\newtheoremstyle{slplain}% name
  {.5\baselineskip\@plus.2\baselineskip\@minus.2\baselineskip}% Space above
  {.5\baselineskip\@plus.2\baselineskip\@minus.2\baselineskip}% Space below
  {\slshape}% Body font
  {}%Indent amount (empty = no indent, \parindent = para indent)
  {\bfseries}%  Thm head font
  {.}%       Punctuation after thm head
  { }%      Space after thm head: " " = normal interword space;
  {}%       Thm head spec

\theoremstyle{slplain}
\newtheorem{rmk}{Remark}

\journal{arXiv}

\begin{document}

\begin{frontmatter}

\title{On the Thermodynamics of the Swift--Hohenberg Theory}

\author{LFR Espath\corref{cor1}\fnref{fn1}}
\ead{espath@gmail.com}

\author{AF Sarmiento\fnref{fn2}}

\author{L Dalcin\fnref{fn2,fn3}}

\author{VM Calo\fnref{fn4,fn5}}

\cortext[cor1]{Corresponding author.}

\fntext[fn1]{Computer, Electrical and Mathematical Sciences and Engineering, King Abdullah University of Science and Technology, Thuwal 23955-6900, Saudi Arabia}

\fntext[fn2]{Extreme Computing Research Center, King Abdullah University of Science and Technology, Thuwal 23955-6900, Saudi Arabia}

\fntext[fn3]{National Scientific and Technical Research Council (CONICET), Sta Fe, Argentina}

\fntext[fn4]{Applied Geology, Western Australian School of Mines, Faculty of Science and Engineering, Curtin University, Perth, WA, Australia 6845}

\fntext[fn5]{Mineral Resources, Commonwealth Scientific and Industrial Research Organisation (CSIRO), Kensington, WA, Australia 6152}

\address{King Abdullah University of Science and Technology (KAUST), Thuwal, Saudi Arabia, and \\
         Curtin University, Bentley, Perth, Western Australia, Australia}

\begin{abstract}
We present the microbalance including the microforces, the first- and second-order microstresses for the Swift--Hohenberg equation concomitantly with their constitutive equations, which are consistent with the free-energy imbalance. We provide an explicit form for the microstress structure for a free-energy functional endowed with second-order spatial derivatives. Additionally, we generalize the Swift--Hohenberg theory via a proper constitutive process. Finally, we present one highly-resolved three-dimensional numerical simulation to demonstrate the particular form of the resulting microstresses and their interactions in the evolution of the Swift--Hohenberg equation.
\end{abstract}

\begin{keyword}
Swift--Hohenberg Theory \sep Thermodynamics of Continua
\end{keyword}

\end{frontmatter}

%===============================================================================================================%

\section{Introduction}

The work of J. Swift and P. C. Hohenberg sought to explain the Rayleigh–Benard instabilities and their patterns (cf. \cite{SWI77}). These well-ordered structures resemble crystalline structures found in material sciences. In their work, entitled ``Hydrodynamic fluctuations at the convective instability'' in 1977, Swift and Hohenberg proposed a new free-energy (cf. equation (21) therein) and a governing equation (cf. equation (19) therein) that motivate our study on the thermodynamics and the microbalance between the microforces and microstresses for this phenomenon as described by the Swift--Hohenberg theory.

Phase-field theories based on first-order gradients were formalized throughout a balance of microforces by \cite{FRI93,FRI94,FRI96,GUR96a}. They sought to segregate the constitutive equations from the balance laws. This dissociation is natural since the balance laws are related to different phenomena whereas the constitutive equations are concerned with particular material behaviors.

Another way to tackle the problem is throughout the virtual power machinery. Some authors investigated its use in other contexts, particularly in solid mechanics, such as \cite{FRI06a} (cf. equations (2-3) therein) and \cite{DEL11}, (cf. equations (55-57) therein). The virtual power machinery was originally proposed by \cite{TOU62,TOU64} to develop a second-order gradient theory.

The Swift--Hohenberg equation can be interpreted as a second-order gradient phase-field model, which broadens its applicability. In the context of solid mechanics, \cite{MIE16} used a phase-field crystal model of ductile fracture in elasto-plastic solids under large strains, where the phase-field approximates sharp cracks. Whereas in the context of fluid mechanics, \cite{PRA15} coupled phase-field crystal with Navier--Stokes to model colloidal suspensions.

In this work, we present a phase-field theory based on second-order gradients. Particularly, we present the thermodynamics and the derivation of the Swift--Hohenberg theory based on a microbalance between the microforce and the first- and second-order microstresses. Moreover, we build a constitutive process for the generalization of the Swift--Hohenberg theory. Finally, to study the interplay between the first- and second-order microstresses, we use highly-resolved simulation and focus on the relevant details of the microstress interactions.

%===============================================================================================================%

\section{Governing Equation}\label{sec:gov.eqs}

Consider the following evolutionary partial differential equation
\begin{equation}\label{eq:pde.dimensional}
\dot{\phi} = - \kappa \eta,
\end{equation}
where $ {\phi = \phi (\T{x}{1}, t)} $ is the order-parameter with $ {\T{x}{1}} $ and $ {t} $ being, respectively, the spatial and temporal coordinates, $ {\kappa > 0} $ is the mass growth coefficient (a constitutive modulus), and $ {\eta} $ is the chemical potential.

\textbf{Notation}: the differential operators $ {\gradx{(\cdot)}} $, $ {\divx{(\cdot)}} $, and $ {\lapx{(\cdot)}} $ represent the gradient, divergence, and Laplacian, respectively. Also, $ {\gradx{\!^2(\cdot)} = \gradx{\gradx{(\cdot)}}} $, $ {\divx{\!^2(\cdot)} = \divx{\divx{(\cdot)}}} $, and $ {\lapx{\!^2(\cdot)} = \lapx{\lapx{(\cdot)}}} $. The symbol $ {\otimes} $ denotes the tensor (dyadic) product. The number of underlines indicates the order of the tensor. The material (or total) time derivative is denoted by $ {()\dot{}} $. Finally, $ {\spart{}{a}{m}} $ is $ {\partn{}{a}{m}} $.

%===============================================================================================================%

\section{Free-Energy Functional}\label{sec:free.energy}

Consider the following class of free-energy functionals
\begin{equation}\label{eq:free.energy}
\begin{split}
\Psi[\phi, \gradx{\phi}, \gradx{\!^2\phi}] = & \int _{\Omega} \psi[\phi, \gradx{\phi}, \gradx{\!^2\phi}] \, \diffo{\Omega} \\
= & \int _{\Omega} \left[ \sum _a \varsigma _a \phi ^a + \gamma \, \gradx{\phi} \cdot \gradx{\phi} + \beta (\T{1}{2} : \gradx{\!^2\phi}) ^2 \right] \diffo{\Omega},
\end{split}
\end{equation}
where $ {\varsigma _a} $, $ {\beta} $, and $ {\gamma} $ are coefficients which control the relative contribution of each term to the energy granting these terms physical meaning. $ {\Omega} $ is a body that occupies a fixed region in a three-dimensional Euclidean space and $ {\Gamma} $ its boundary. Here, $ {\gradx{\!^2\phi} = \gradx{\gradx{\phi}}} $ is the Hessian of the order-parameter $ {\phi} $ and $ {\T{1}{2} = \T{e}{1} _i \otimes \T{e}{1} _j} $ is the second-order identity tensor, where the set $ {\T{e}{1} _i} $ (with $ {i=1,2,} $ and $ {3} $) forms an orthonormal Cartesian basis.

The equations that define the free energy (\ref{eq:free.energy}) together with (\ref{eq:pde.dimensional}) yield the classical Swift--Hohenberg equation (cf. \cite{SWI77}).

\begin{rmk}[Functional derivative]\label{rmk:functional.derivative}
Consider the functional in the form (\ref{eq:free.energy}), i.e., \emph{$ {\Psi[\phi, \gradx{\phi}, \gradx{\!^2\phi}]} $}. Thus, its first variation is
\emph{
\begin{equation}\label{eq:first.variation.1}
\delta \Psi = \int _{\Omega} [\spart{\psi}{\phi}{} \delta \phi + \spart{\psi}{\gradx{\phi}}{} \cdot \delta \gradx{\phi} + \spart{\psi}{\gradx{\!^2\phi}}{} : \delta \gradx{\!^2\phi}] \, \diffo{\Omega}.
\end{equation}
}
Taking into account that the differential and variational operators are commutative, thus, the following identities hold
\emph{
\begin{subequations}\label{eq:functional.identities}
\begin{align}
\divx{(\spart{\psi}{\gradx{\phi}}{} \delta \phi)} = & \, \divx{(\spart{\psi}{\gradx{\phi}}{})} \delta \phi + \spart{\psi}{\gradx{\phi}}{} \cdot \delta \gradx{\phi}, \label{eq:fn.id.1} \\
\divx{(\spart{\psi}{\gradx{\!^2\phi}}{} \cdot \delta \gradx{\phi})} = & \, \divx{\spart{\psi}{\gradx{\!^2\phi}}{}} \cdot \delta \gradx{\phi} + \spart{\psi}{\gradx{\!^2\phi}}{} : \delta \gradx{\!^2\phi}, \label{eq:fn.id.2} \\
\divx{(\divx{(\spart{\psi}{\gradx{\!^2\phi}}{})} \delta \phi)} = & \, \divx{(\divx{\spart{\psi}{\gradx{\!^2\phi}}{}})} \delta \phi + \divx{\spart{\psi}{\gradx{\!^2\phi}}{}} \cdot \delta \gradx{\phi}, \label{eq:fn.id.3}
\end{align}
\end{subequations}
}
from (\ref{eq:first.variation.1}), we obtain
\emph{
\begin{equation}\label{eq:first.variation.2}
\begin{split}
\delta \Psi = & \int _{\Omega} \left[\spart{\psi}{\phi}{} - \divx{\spart{\psi}{\gradx{\phi}}{}} + \divx{\!^2\spart{\psi}{\gradx{\!^2\phi}}{}} \right] \delta \phi \, \diffo{\Omega} \\
& + \int _{\Gamma} \left[(\spart{\psi}{\gradx{\phi}}{} - \divx{\spart{\psi}{\gradx{\!^2\phi}}{}}) \delta \phi + \spart{\psi}{\gradx{\!^2\phi}}{} \cdot \delta \gradx{\phi}\right] \cdot \T{n}{1} \, \diffo{\Gamma},
\end{split}
\end{equation}
}
where $ {\T{n}{1}} $ is the unit, outward normal to the domain $ {\Omega} $.
\qed
\end{rmk}

Finally, the chemical potential $ {\eta} $ is
\begin{equation}\label{eq:chemical.potential}
\eta = \dfrac{\delta \Psi}{\delta \phi} = \spart{\psi}{\phi}{} - \divx{\spart{\psi}{\gradx{\phi}}{}} + \divx{\!^2\spart{\psi}{\gradx{\!^2\phi}}{}}.
\end{equation}

We use different expressions of this identity $ {\lapx{\phi} = \tr{\gradx{\!^2\phi}} = \T{1}{2} : \gradx{\!^2\phi}} $ according to what simplifies the derivations most.

\begin{rmk}[Chemical potential for the free energy (\ref{eq:free.energy})]\label{rmk:chemical.potential}
The components of the functional derivative are
\emph{
\begin{subequations}\label{eq:ch.1a}
\begin{align}
\spart{\psi}{\phi}{} = & \, \textstyle{\sum _a} a \varsigma \phi ^{(a-1)}, \label{eq:ch.1a.01} \\
\divx{\spart{\psi}{\gradx{\phi}}{}} = & \, \divx{[2 \gamma \T{1}{2} \cdot \gradx{\phi}]} = 2 \gamma \lapx{\phi}, \label{eq:ch.1.02} \\
\divx{\!^2\spart{\psi}{\gradx{\!^2\phi}}{}} = & \, \divx{\!^2[2 \beta (\T{1}{2} : \gradx{\!^2\phi}) \T{1}{2}]} = 2 \beta \lapx{\!^2\phi}. \label{eq:ch.1a.02}
\end{align}
\end{subequations}
}
Then, the chemical potential reads
\emph{
\begin{equation}\label{eq:ch.1b}
\eta =  \sum _a a \varsigma \phi ^{(a-1)} - 2 \gamma \lapx{\phi} + 2 \beta \lapx{\!^2\phi}.
\end{equation}
}
\qed
\end{rmk}

%===============================================================================================================%

\section{Microbalance and Free-Energy Imbalance}\label{sec:micro.balance}

To start, we account for the rate of work done by external agencies on each kinematic process -- i.e., \emph{external agencies} $ {\times} $ \emph{kinematic processes} -- to build the free-energy imbalance since there exist energies in this physical law governed by the Swift--Hohenberg equation. In phase-field theories, the kinematic processes are related to the order-parameter $ {\phi} $. We thus must account for rate of work stated on the temporal changes in the order-parameter and its gradients. Finally, considering that the free-energy (\ref{eq:free.energy}) is endowed with up to second-order gradients, the external chemical power expenditure $ {\dot{w} _c ^{ext}} $ has the following form
\begin{equation}\label{eq:external.chemical.power}
\dot{w} _c ^{ext} = \divx{\left[\dot{\phi} (\T{\xi}{1} - \divx{\T{\Xi}{2}}) + \gradx{\dot{\phi}} \cdot \T{\Xi}{2} \right]} + \dot{\phi} \varpi,
\end{equation}
being $ {\T{\xi}{1}} $ and $ {\T{\Xi}{2}} $ the first- and second-order microstresses -- stresses-like objects -- respectively, and $ {\varpi} $ a scalar body microforce that represents the external force. The choice of an external chemical power expenditure in the form of (\ref{eq:external.chemical.power}) is supported by an internal chemical power expenditure -- detailed in remark \ref{rmk:conjugate.pairs}, cf. (\ref{eq:powers.domain}) -- as expected in second-order theories. Additionally, to obtain a proper set of conjugate pairs stated on remark \ref{rmk:conjugate.pairs}, the local microbalance assumes the following new form
\begin{equation}\label{eq:microbalance.1}
\divx{(\T{\xi}{1} - \divx{\T{\Xi}{2}})} + \pi + \varpi = 0,
\end{equation}
where $ {\pi} $ is scalar body microforce that represents the internal microforce.

\begin{rmk}[Conjugate Pairs]\label{rmk:conjugate.pairs}
Consider the external chemical power expenditure (\ref{eq:external.chemical.power}) in its integral form
\emph{
\begin{equation}\label{eq:powers.boundaries}
\int _{\Omega} \dot{\phi} \varpi \, \diffo{\Omega} + \int _{\Gamma} [\dot{\phi} (\T{\xi}{1} - \divx{\T{\Xi}{2}})] \cdot \T{n}{1} \, \diffo{\Gamma} + \int _{\Gamma} [\gradx{\dot{\phi}} \cdot \T{\Xi}{2}] \cdot \T{n}{1} \, \diffo{\Gamma},
\end{equation}
}
which states that there are two traction-like objects, a zeroth-order effective microtraction \emph{$ {(\T{\xi}{1} - \divx{\T{\Xi}{2}}) \cdot \T{n}{1}} $} and first-order microtraction \emph{$ {\T{\Xi}{2} \cdot \T{n}{1}} $}. Although \emph{$ {\dot{\phi}} $} and \emph{$ {\gradx{\dot{\phi}}} $} are not independent kinematic processes -- i.e., \emph{$ {\dot{\phi}} $} and \emph{$ {\gradx{\dot{\phi}}} $} cannot be prescribed independently -- the external power may be rewritten, which is rigorously equivalent to (\ref{eq:powers.boundaries}), as two new zeroth-order microtractions that are power-conjugate to two independent kinematic processes, $ {\dot{\phi}} $ and $ {\spart{\dot{\phi}}{n}{}} $ (cf. \cite{FRI06a}).

Now, using the divergence theorem in (\ref{eq:powers.boundaries}) and the microbalance equation (\ref{eq:microbalance.1}), we obtain the internal chemical power expenditure in its integral form,
\emph{
\begin{equation}\label{eq:powers.domain}
- \int _{\Omega} \dot{\phi} \pi \, \diffo{\Omega} + \int _{\Omega} \gradx{\dot{\phi}} \cdot \T{\xi}{1} \, \diffo{\Omega} + \int _{\Omega} \gradx{\!^2\dot{\phi}} : \T{\Xi}{2} \, \diffo{\Omega},
\end{equation}
}
depicting three power-conjugate pairs: \emph{$ {(\dot{\phi}, -\pi)} $}, \emph{$ {(\gradx{\dot{\phi}}, \T{\xi}{1})} $}, and \emph{$ {(\gradx{\!^2\dot{\phi}}, \T{\Xi}{2})} $}.
\qed
\end{rmk}
Since $ {\gradx{\!^2\dot{\phi}}} $ is symmetric, we assume that $ {\T{\Xi}{2}} $ is as well, without loss of generality.

The free-energy imbalance states that the rate at which the free-energy changes in time has an upper bound given by the external rate of work, i.e.,
\begin{equation}\label{eq:free.energy.imbalance.1}
\dot{\psi} \leqslant \divx{\left[\dot{\phi} (\T{\xi}{1} - \divx{\T{\Xi}{2}}) + \gradx{\dot{\phi}} \cdot \T{\Xi}{2} \right]} + \dot{\phi} \varpi.
\end{equation}

Here, we apply the Coleman--Noll procedure (cf. \cite{COL63}) considering the following list of variables and their dependencies
\begin{equation}\label{eq:dependencies.1}
\{\psi, \T{\xi}{1}, \T{\Xi}{2}, \pi\} = f(\phi, \gradx{\phi}, \gradx{\!^2\phi}).
\end{equation}
The set of functions $ {\{\psi, \T{\xi}{1}, \T{\Xi}{2}, \pi\}} $ is called \emph{thermodynamic} or \emph{constitutive process} if the conservation laws, microbalance and energy balance in our case, are satisfied (cf. §2, \cite{COL63}). This process is called \emph{admissible} if it obeys the local free-energy imbalance (or Clausius--Duhem inequality) and is endowed with a positive-definite finite absolute temperature. 

Now, considering the explicit dependencies listed in (\ref{eq:dependencies.1}) and substituting them into the free-energy imbalance (\ref{eq:free.energy.imbalance.1}), we obtain
\begin{equation}\label{eq:free.energy.imbalance.2}
\begin{split}
\spart{\psi}{\phi}{} \dot{\phi} + \spart{\psi}{\gradx{\phi}}{} \cdot (\gradx{\phi}) \dot{} + \spart{\psi}{\gradx{\!^2\phi}}{} : (\gradx{\!^2\phi}) \dot{} \leqslant & \, \gradx{\dot{\phi}} \cdot \T{\xi}{1} + \dot{\phi} \, \divx{\T{\xi}{1}} - \gradx{\dot{\phi}} \cdot \divx{\T{\Xi}{2}} \\
& - \dot{\phi} \, \divx{\!^2\T{\Xi}{2}} + \gradx{\!^2\dot{\phi}} : \T{\Xi}{2} + \gradx{\dot{\phi}} \cdot \divx{\T{\Xi}{2}} + \dot{\phi} \varpi.
\end{split}
\end{equation}
Without constraints, the constitutive relations (\ref{eq:dependencies.1}) might violate the free-energy imbalance (\ref{eq:free.energy.imbalance.2}). Rearranging equation (\ref{eq:free.energy.imbalance.2}) and considering that there is no advection, the spatial and temporal derivatives commute. We thus build a \emph{constitutive process} for $ {\psi} $, $ {\pi} $, $ {\T{\xi}{1}} $, and $ {\T{\Xi}{2}} $ which implies that
\begin{equation}\label{eq:free.energy.imbalance.3}
(\spart{\psi}{\phi}{} \underbrace{- \varpi - \divx{\T{\xi}{1}} + \divx{\!^2\T{\Xi}{2}}}_{\pi}) \dot{\phi} + (\spart{\psi}{\gradx{\phi}}{} - \T{\xi}{1}) \cdot \gradx{\dot{\phi}} + (\spart{\psi}{\gradx{\!^2\phi}}{} - \T{\Xi}{2}) : \gradx{\!^2\dot{\phi}} \leqslant 0.
\end{equation}
At some chosen point $ {(\T{x}{1}, t)} $ we can set a field $ {\phi} $ such that $ {\dot{\phi}} $, $ {\gradx{\phi}} $, $ {\gradx{\dot{\phi}}} $, $ {\gradx{\!^2\phi}} $, and $ {\gradx{\!^2\dot{\phi}}} $ have arbitrary values. Thus, one has to find the proper constitutive relations for $ {\{\psi, \T{\xi}{1}, \T{\Xi}{2}, \pi\}} $ to guarantee the inequality direction. Thus, taking into account that (\ref{eq:free.energy.imbalance.3}) is linear on $ {\dot{\phi}} $, $ {\gradx{\dot{\phi}}} $, and $ {\gradx{\!^2\dot{\phi}}} $, only the trivial solutions exist. Therefore, the constitutive relations for $ {\T{\xi}{1}} $, $ {\T{\Xi}{2}} $, and $ {\pi} $ are
\begin{subequations}\label{eq:micro.constitutive.1}
\begin{align}
\pi = & \, - \spart{\psi}{\phi}{} = - \textstyle{\sum _a} a \varsigma \phi ^{(a-1)}, \label{eq:micro.const.1.1} \\
\T{\xi}{1} = & \, \spart{\psi}{\gradx{\phi}}{} = 2 \gamma \, \gradx{\phi}, \label{eq:micro.const.1.2} \\
\T{\Xi}{2} = & \, \spart{\psi}{\gradx{\!^2\phi}}{} = 2 \beta \lapx{\phi} \, \T{1}{2}, \label{eq:micro.const.1.3}
\end{align}
\end{subequations}
where the identities listed in (\ref{eq:ch.1a}) where used to express these relations explicitly. Here, we observe that our microbalance (\ref{eq:microbalance.1}) with its constitutive relations (\ref{eq:micro.constitutive.1}) recovers the steady state of the Swift--Hohenberg equation, i.e., $ {\frac{\delta \Psi}{\delta \phi} = 0} $.

To recover the Swift--Hohenberg equation as it was conceived, we have to include  $ {\dot{\phi}} $ in the list of variables. Thus, the new list is
\begin{equation}\label{eq:dependencies.2}
\{\psi, \T{\xi}{1}, \T{\Xi}{2}, \pi\} = f(\phi, \gradx{\phi}, \gradx{\!^2\phi}, \dot{\phi}),
\end{equation}
and its free-energy imbalance
\begin{equation}\label{eq:free.energy.imbalance.4}
(\spart{\psi}{\phi}{} + \pi) \dot{\phi} + (\spart{\psi}{\gradx{\phi}}{} - \T{\xi}{1}) \cdot \gradx{\dot{\phi}} + (\spart{\psi}{\gradx{\!^2\phi}}{} - \T{\Xi}{2}) : \gradx{\!^2\dot{\phi}} + \spart{\psi}{\dot{\phi}}{} \ddot{\phi} \leqslant 0.
\end{equation}
Now, the inequality (\ref{eq:free.energy.imbalance.4}) is no longer linear on $ {\dot{\phi}} $, admitting nontrivial solutions in its first term. Thus, we define $ {\pi_{\text{dis}} = \pi + \spart{\psi}{\phi}{}} $. The last terms admit only trivial solutions, therefore $ {\spart{\psi}{\dot{\phi}}{} = 0} $.

The thermodynamic constraints yield the consistent constitutive relations
\begin{subequations}\label{eq:micro.constitutive.2}
\begin{align}
\pi_{\text{dis}} \dot{\phi} = & \, (\pi + \spart{\psi}{\phi}{}) \dot{\phi} \leqslant 0, \label{eq:micro.const.2.1} \\
\T{\xi}{1} = & \, \spart{\psi}{\gradx{\phi}}{}, \label{eq:micro.const.2.2} \\
\T{\Xi}{2} = & \, \spart{\psi}{\gradx{\!^2\phi}}{}. \label{eq:micro.const.2.3}
\end{align}
\end{subequations}
From (\ref{eq:micro.const.2.1}), one has several choices; however, to recover the Swift--Hohenberg equation we choose $ {\pi_{\text{dis}} = - \frac{\dot{\phi}}{\kappa}} $, being $ {\kappa > 0} $ the constitutive modulus of this equation, and set $ {\varpi = 0} $.

%===============================================================================================================%

\section{First and Second Laws of Thermodynamics}\label{sec:thermodynamics}

Here, we detail the derivation of the first and second laws of thermodynamics in the usual manner, i.e., directly from the governing equation and the free-energy (cf. \cite{GUR10}), to further validate our model for the microbalance.

\subsection*{\centering First Law}

The product between the material derivative of the phase-field and the chemical potential ($ {\dot{\phi} \eta} $) yields the balance between the external and internal chemical rate of work.

Consider the free energy in the form (\ref{eq:free.energy}). In the general case,
\begin{equation}\label{eq:general.dot.psi}
\dot{\psi} = \spart{\psi}{\phi}{} \, \dot{\phi} + \spart{\psi}{\gradx{\phi}}{} \cdot (\gradx{\phi}) \dot{} + \spart{\psi}{\gradx{\!^2\phi}}{} : (\gradx{\!^2\phi}) \dot{} + \spart{}{\theta}{} \psi \, \dot{\theta},
\end{equation}
where $ {\theta} $ is the absolute temperature, whereas an isothermal process implies
\begin{equation}\label{eq:dot.psi}
\dot{\psi} = \spart{\psi}{\phi}{} \, \dot{\phi} + \spart{\psi}{\gradx{\phi}}{} \cdot (\gradx{\phi}) \dot{} + \spart{\psi}{\gradx{\!^2\phi}}{} : (\gradx{\!^2\phi}) \dot{}.
\end{equation}

% We seek to express the energy balance in the following form
% \begin{equation}\label{eq:power.balance}
% \spart{\psi}{\phi}{} \, \dot{\phi} + \spart{}{\gradx{\phi}}{} \psi \cdot (\gradx{\phi}) \dot{} + \spart{\psi}{\gradx{\!^2\phi}}{} : (\gradx{\!^2\phi}) \dot{} + \dot{w} _c ^{int} = \dot{w} _c ^{ext},
% \end{equation}
% where $ {\dot{w} _c ^{int}} $ and $ {\dot{w} _c ^{ext}} $ account for the internal and external chemical powers, respectively. Thus, in an isothermal evolution of the system, equation (\ref{eq:power.balance}) reads $ {\dot{\psi} + \dot{w} _c ^{int} = \dot{w} _c ^{ext}} $.

\begin{rmk}[Identities]\label{rmk:identities}
The identities used in this section are
\emph{
\begin{subequations}\label{eq:identities}
\begin{align}
\divx{(\dot{\phi} \, \spart{\psi}{\gradx{\phi}}{})} = & \, \gradx{\dot{\phi}} \cdot \spart{\psi}{\gradx{\phi}}{} + \dot{\phi} \, \divx{\spart{\psi}{\gradx{\phi}}{}}, \label{id:1} \\
\divx{(\dot{\phi} \, \divx{\spart{\psi}{\gradx{\!^2\phi}}{}})} = & \, \gradx{\dot{\phi}} \cdot \divx{\spart{\psi}{\gradx{\!^2\phi}}{}} + \dot{\phi} \, \divx{\!^2\spart{\psi}{\gradx{\!^2\phi}}{}}, \label{id:2} \\
\divx{(\gradx{\dot{\phi}} \cdot \spart{\psi}{\gradx{\!^2\phi}}{})} = & \, \gradx{\!^2\dot{\phi}} : \spart{\psi}{\gradx{\!^2\phi}}{} + \gradx{\dot{\phi}} \cdot \divx{\spart{\psi}{\gradx{\!^2\phi}}{}}. \label{id:3}
\end{align}
\end{subequations}
}
\qed
\end{rmk}

Here, we analyze the chemical powers using $ {\dot{\phi} \eta} $, i.e.,
\begin{equation}\label{eq:chemical.powers.01}
\dot{\phi} \eta = \dot{\phi} (\spart{\psi}{\phi}{} - \divx{\spart{\psi}{\gradx{\phi}}{}} + \divx{\!^2\spart{\psi}{\gradx{\!^2\phi}}{}}).
\end{equation}

Rewriting equation (\ref{eq:chemical.powers.01}) using the identities (\ref{eq:identities}), we obtain
\begin{equation}\label{eq:chemical.powers.02}
\begin{split}
\dot{\phi} \eta = & \, \spart{}{\phi}{} \psi \, \dot{\phi} + \spart{}{\gradx{\phi}}{} \psi \cdot \gradx{\dot{\phi}} + \spart{\psi}{\gradx{\!^2\phi}}{} : \gradx{\!^2\dot{\phi}} \\
& \, - \divx{\left(\dot{\phi} \, \spart{\psi}{\gradx{\phi}}{} - \dot{\phi} \, \divx{\spart{\psi}{\gradx{\!^2\phi}}{}} + \gradx{\dot{\phi}} \cdot \spart{\psi}{\gradx{\!^2\phi}}{}\right)}.
\end{split}
\end{equation}

Considering that there is no advection, the spatial and temporal derivatives commute, and using equations (\ref{eq:pde.dimensional}) and (\ref{eq:chemical.powers.02}), we obtain
\begin{multline}\label{eq:chemical.powers.03}
\spart{\psi}{\phi}{} \, \dot{\phi} + \spart{}{\gradx{\phi}}{} \psi \cdot (\gradx{\phi}) \dot{} + \spart{\psi}{\gradx{\!^2\phi}}{} : (\gradx{\!^2\phi}) \dot{} + \kappa \eta ^2  = \\
\divx{\left(\dot{\phi} \, \spart{\psi}{\gradx{\phi}}{} - \dot{\phi} \, \divx{\spart{\psi}{\gradx{\!^2\phi}}{}} + \gradx{\dot{\phi}} \cdot \spart{\psi}{\gradx{\!^2\phi}}{}\right)},
\end{multline}
alternatively, if no external power is spent across the boundaries
\begin{equation}\label{eq:chemical.powers.04}
\dot{\psi} = - \kappa \eta ^2.
\end{equation}

The internal and external powers are
\begin{subequations}\label{eq:powers.01}
\begin{align}
\dot{w} _c ^{int} = & \, \kappa \eta ^2 + \spart{\psi}{\phi}{} \, \dot{\phi} + \spart{}{\gradx{\phi}}{} \psi \cdot (\gradx{\phi}) \dot{} + \spart{\psi}{\gradx{\!^2\phi}}{} : (\gradx{\!^2\phi}) \dot{}, \label{eq:powers.int.c.1} \\
\dot{w} _c ^{ext} = & \, \divx{\left(\dot{\phi} \, \spart{\psi}{\gradx{\phi}}{} - \dot{\phi} \, \divx{\spart{\psi}{\gradx{\!^2\phi}}{}} + \gradx{\dot{\phi}} \cdot \spart{\psi}{\gradx{\!^2\phi}}{}\right)}. \label{eq:powers.ext.c.1}
\end{align}
\end{subequations}
Note that $ {\kappa \eta ^2 + \spart{\psi}{\phi}{} \, \dot{\phi} = - \dot{\phi} \pi} $.

The first law of thermodynamics represents the energy balance in the system and states explicitly the interplay between the kinetic energy $ {e_k} $, the internal energy $ {e_i} $, the rate at which (mechanical and chemical) power is spent, and the rate at which energy in the form of heat is transferred, i.e.,
\begin{equation}
\dot{e_T} = \dot{e} _k + \dot{e} _i = \dot{w} _m ^{ext} + \dot{w} _c ^{ext} - \divx{\T{q}{1}} + q,
\end{equation}
where $ {\T{q}{1}} $ is the heat flux, and $ {q} $ may either be a heat sink or source.

Finally, we obtain the first law of thermodynamics
\begin{equation}\label{eq:first.law}
\begin{split}
\dot{e} _i = \spart{}{\phi}{} \psi \, \dot{\phi} + \spart{}{\gradx{\phi}}{} \psi \cdot (\gradx{\phi}) \dot{} + \spart{\psi}{\gradx{\!^2\phi}}{} : (\gradx{\!^2\phi}) \dot{} + \kappa \eta ^2 - \divx{\T{q}{1}} + q,
\end{split}
\end{equation}
In the absence of heat transfer, the first law of thermodynamics (i.e., the balance of energy) reads
\begin{equation}\label{eq:first.law.no.heat}
\begin{split}
\dot{e} _i = \dot{\psi} + \kappa \eta ^2,
\end{split}
\end{equation}
where we used equation (\ref{eq:powers.01}).

\subsection*{\centering Second Law}

The second law of thermodynamics (in the form of the Clausius--Duhem inequality or entropy imbalance) states that the entropy $ {s} $ should grow at least at the rate given by the entropy flux $ {\T{q}{1}/\theta} $ added to the entropy supply $ {q/\theta} $, i.e.,
\begin{equation}\label{eq:entropy.growth}
\dot{s} \geqslant - \divx{\left(\dfrac{\T{q}{1}}{\theta}\right)} + \dfrac{q}{\theta} = \dfrac{1}{\theta} \left(- \divx{\T{q}{1} + \dfrac{1}{\theta} \gradx{\theta} \cdot \T{q}{1}} + q \right).
\end{equation}
By definition, the free energy is
\begin{equation}
\psi = e _i - \theta s.
\end{equation}
Taking the material time derivative, we obtain
\begin{equation}\label{eq:free.energy.time}
\dot{\psi} = \dot{e} _i - \dot{\theta} s - \theta \dot{s}.
\end{equation}

Substituting the first law (\ref{eq:first.law}) and equation (\ref{eq:free.energy.time}) into equation (\ref{eq:entropy.growth}), and using standard arguments of Coleman and Noll (cf. \cite{COL63}) the entropy is $ {s = - \spart{}{\theta}{} \psi} $, and the second law of thermodynamics is obtained in the form of the entropy imbalance, i.e.,
\begin{equation}\label{eq:entropy.imbalance}
\dot{s} = \dfrac{1}{\theta} \left( \kappa \eta ^2 - \dfrac{1}{\theta} \gradx{\theta} \cdot \T{q}{1} \right) \geqslant 0,
\end{equation}
In the absence of heat transfer, the entropy imbalance yields
\begin{equation}\label{eq:entropy.imbalance.no.heat}
\kappa \eta ^2 \geqslant 0.
\end{equation}
Finally, we conclude from (\ref{eq:entropy.imbalance.no.heat}) and (\ref{eq:chemical.powers.04}) that $ {\dot{\psi} \leqslant 0} $ if no external power is spent across the boundaries.

%===============================================================================================================%

\section{A \emph{Constitutive Process} for the Generalization of the Swift--Hohenberg Theory}\label{sec:general.constitutive.process}

Here, we generalize the dependencies in the list of variables (\ref{eq:dependencies.1}) and define
\begin{equation}\label{eq:general.dependencies}
\{\psi, \T{\xi}{1}, \T{\Xi}{2}, \pi\} = f(\phi, \gradx{\phi}, \gradx{\!^2\phi}, \dot{\phi}, \gradx{\dot{\phi}}, \gradx{\!^2\dot{\phi}}).
\end{equation}
The free-energy imbalance is now given by
\begin{multline}\label{eq:free.energy.imbalance.5}
(\spart{\psi}{\phi}{} + \pi) \dot{\phi} + (\spart{\psi}{\gradx{\phi}}{} - \T{\xi}{1}) \cdot \gradx{\dot{\phi}} + (\spart{\psi}{\gradx{\!^2\phi}}{} - \T{\Xi}{2}) : \gradx{\!^2\dot{\phi}} \\
+ \spart{\psi}{\dot{\phi}}{} \ddot{\phi} + \spart{\psi}{\gradx{\dot{\phi}}}{} \cdot \gradx{\ddot{\phi}} + \spart{\psi}{\gradx{\!^2\dot{\phi}}}{} : \gradx{\!^2\ddot{\phi}} \leqslant 0.
\end{multline}
While the inequality is no longer linear on $ {\dot{\phi}} $, $ {\gradx{\dot{\phi}}} $, and $ {\gradx{\!^2\dot{\phi}}} $, admitting nontrivial solutions for its first three terms, the last three terms admit only trivial solutions, i.e., $ {\spart{\psi}{\dot{\phi}}{} = \spart{\psi}{\gradx{\dot{\phi}}}{} = \spart{\psi}{\gradx{\!^2\dot{\phi}}}{} = 0} $. Thus, we define
\begin{subequations}\label{eq:micro.diss}
\begin{align}
\pi_{\text{dis}} = & \, \pi + \spart{\psi}{\phi}{}, \label{eq:micro.diss.1} \\
\T{\xi}{1}_{\text{dis}} = & \, \T{\xi}{1} - \spart{\psi}{\gradx{\phi}}{}, \label{eq:micro.diss.2} \\
\T{\Xi}{2}_{\text{dis}} = & \, \T{\Xi}{2} - \spart{\psi}{\gradx{\!^2\phi}}{}. \label{eq:micro.diss.3}
\end{align}
\end{subequations}
Now, the inequality to be enforced is
\begin{equation}\label{eq:free.energy.imbalance.6}
\pi_{\text{dis}} \dot{\phi} - \T{\xi}{1}_{\text{dis}} \cdot \gradx{\dot{\phi}} - \T{\Xi}{2}_{\text{dis}} : \gradx{\!^2\dot{\phi}} \leqslant 0.
\end{equation}

For simplicity, we assume that $ {\pi_{\text{dis}}} $, $ {\gradx{\dot{\phi}}} $, and $ {\T{\Xi}{2}_{\text{dis}}} $ are defined as a linear combination of $ {\dot{\phi}} $, $ {\gradx{\dot{\phi}}} $, and $ {\gradx{\!^2\dot{\phi}}} $. Thus,
\begin{subequations}\label{eq:linear.micro.diss}
\begin{align}
\pi_{\text{dis}} = & \, - \alpha \dot{\phi} - \T{a}{1} \cdot \gradx{\dot{\phi}} - \T{A}{2} : \gradx{\!^2\dot{\phi}}, \label{eq:linear.micro.diss.1} \\
\T{\xi}{1}_{\text{dis}} = & \, \T{\sigma}{1} \dot{\phi} + \T{S}{2} \cdot \gradx{\dot{\phi}} + \T{\Sigma}{3} : \gradx{\!^2\dot{\phi}}, \label{eq:linear.micro.diss.2} \\
\T{\Xi}{2}_{\text{dis}} = & \, \T{U}{2} \dot{\phi} + \gradx{\dot{\phi}} \cdot \T{\Upsilon}{3} + \T{G}{4} : \gradx{\!^2\dot{\phi}}. \label{eq:linear.micro.diss.3}
\end{align}
\end{subequations}
Rearranging (\ref{eq:free.energy.imbalance.6}) and considering (\ref{eq:linear.micro.diss}), we obtain
\begin{multline}\label{eq:free.energy.imbalance.7}
- \alpha \dot{\phi} ^2 - \T{S}{2} : \gradx{\dot{\phi}} \otimes \gradx{\dot{\phi}} - \T{G}{4} :: \gradx{\!^2\dot{\phi}} \otimes \gradx{\!^2\dot{\phi}} \\
- (\T{a}{1} + \T{\sigma}{1}) \cdot (\dot{\phi} \, \gradx{\dot{\phi}}) - (\T{A}{2} + \T{U}{2}) : (\dot{\phi} \, \gradx{\!^2\dot{\phi}}) - (\T{\Upsilon}{3} + \T{\Sigma}{3}) \threevdots \gradx{\!^2\dot{\phi}} \otimes \gradx{\dot{\phi}} \leqslant 0.
\end{multline}
We assume that we cannot assert the direction of the inequality (\ref{eq:free.energy.imbalance.7}) for all terms, but we can state that $ {\alpha} $, $ {\T{S}{2}} $, and $ {\T{G}{4}} $ must be positive definite. Additionally, assuming that $ {\{\alpha, \T{a}{1}, \T{A}{2}, \T{\sigma}{1}, \T{S}{2}, \T{\Sigma}{3}, \T{U}{2}, \T{\Upsilon}{3}, \T{G}{4}\}} $ are constant coefficients, we can look for the trivial solution for the remaining terms. Thus, assuming that $ {\T{a}{1} + \T{\sigma}{1} = 0} $, $ {\T{A}{2} + \T{U}{2} = 0} $, and $ {\T{\Upsilon}{3} + \T{\Sigma}{3} = 0} $, we obtain
\begin{subequations}\label{eq:linear.micro}
\begin{align}
\pi = & \, - \spart{\psi}{\phi}{} - \alpha \dot{\phi} - \T{a}{1} \cdot \gradx{\dot{\phi}} - \T{A}{2} : \gradx{\!^2\dot{\phi}}, \label{eq:linear.micro.1} \\
\T{\xi}{1} = & \, \spart{\psi}{\gradx{\phi}}{} - \dot{\phi} \T{a}{1} + \T{S}{2} \cdot \gradx{\dot{\phi}} + \T{\Sigma}{3} : \gradx{\!^2\dot{\phi}}, \label{eq:linear.micro.2} \\
\T{\Xi}{2} = & \, \spart{\psi}{\gradx{\!^2\phi}}{} - \T{A}{2} \dot{\phi} - \gradx{\dot{\phi}} \cdot \T{\Sigma}{3} + \T{G}{4} : \gradx{\!^2\dot{\phi}}. \label{eq:linear.micro.3}
\end{align}
\end{subequations}

Finally, using the constitutive relations (\ref{eq:linear.micro}) in the new form of the microbalance (\ref{eq:microbalance.1}), we obtain the generalized Swift--Hohenberg equation
\begin{multline}\label{eq:generalization.SH.1}
\divx{\left[\spart{\psi}{\gradx{\phi}}{} - \dot{\phi} \T{a}{1} + \T{S}{2} \cdot \gradx{\dot{\phi}} + \T{\Sigma}{3} : \gradx{\!^2\dot{\phi}} - \divx{\left(\spart{\psi}{\gradx{\!^2\phi}}{} - \T{A}{2} \dot{\phi} - \gradx{\dot{\phi}} \cdot \T{\Sigma}{3} + \T{G}{4} : \gradx{\!^2\dot{\phi}}\right)}\right]} \\
- \spart{\psi}{\phi}{} - \alpha \dot{\phi} - \T{a}{1} \cdot \gradx{\dot{\phi}} - \T{A}{2} : \gradx{\!^2\dot{\phi}} + \varpi = 0,
\end{multline}
and considering the free-energy (\ref{eq:free.energy}), we obtain
\begin{multline}\label{eq:generalization.SH.2}
- \textstyle{\sum _a} a \varsigma \phi ^{(a-1)} + 2 \gamma \, \lapx{\phi} - 2 \beta \, \lapx{\!^2\phi} \\
- \alpha \dot{\phi} - 2 \, \T{a}{1} \cdot \gradx{\dot{\phi}} + \T{S}{2} : \gradx{\!^2\dot{\phi}} + 2 \, \T{\Sigma}{3} \threevdots \gradx{\!^3\dot{\phi}} - \T{G}{4} :: \gradx{\!^4\dot{\phi}} + \varpi = 0.
\end{multline}

%===============================================================================================================%

\section{Numerical Experiment}\label{sec:numerics}

Here, we present a highly-resolved three-dimensional simulation to depict the interplay between the first- and second-order microstresses. We use a high-order NURBS-based finite element solver, PetIGA: \cite{DAL16}, which has been used extensibly in the modeling of multiphyscis processes including phase-field applications, \cite{THI13,SAG15,VIG15a,VIG15b,VIG16a,SAR16}. We solve the resulting equation in its weak primal version, the regular version Swift--Hohenberg equation, composed by equations (\ref{eq:pde.dimensional}) and (\ref{eq:chemical.potential}). We employ a tensor-product B-spline approximation with $ {64^3} $ elements of polynomial degree $ {4} $ with $ {C^3} $ continuity at element interfaces.

The free-energy coefficients in equation (\ref{eq:free.energy}) are given by
\begin{equation}\label{eq:free.energy.coefficients}
\varsigma _1 = 0; \quad \varsigma _2 = - \dfrac{1}{2}; \quad  \varsigma _3 = 0; \quad \varsigma _4 = \dfrac{1}{4}; \quad \varsigma _{a>4} = 0; \quad \gamma = -1; \quad \beta = \dfrac{1}{2},
\end{equation}
while the constitutive modulus is $ {\kappa = 1} $ in (\ref{eq:pde.dimensional}) and the initial condition is defined as
\begin{equation}\label{eq:initial.condition}
\phi(\T{x}{1},0) = \tanh \left[ \dfrac{\left(x_1 - 20 - r_e \cos \left(\dfrac{8 x_3 \pi^{\flat}}{L}\right)\right)^2 + \left(x_2 - 20 - r_e \sin \left(\dfrac{10 x_3 \pi^{\flat}}{L}\right)\right)^2 - r_c^2}{t_h} \right],
\end{equation}
where $ {\pi^{\flat} = 3.14159 \ldots} $, $ {L=40} $, $ {r_e = 0.02 L} $, $ {r_c = 0.075 L} $, and $ {t_h = 0.0625 L} $. The domain is a cube with dimensions $ {(L_{x_1}, L_{x_2}, L_{x_3}=L)} $, with periodic boundary conditions.

The term $ {\sum _a \varsigma _a \phi ^a} $ in the free-energy (\ref{eq:free.energy}) is a double-well potential function with the coefficients listed in (\ref{eq:free.energy.coefficients}). This function is defined in $ {[-\sqrt{2},\sqrt{2}]} $ with minima at $ {-1} $ and $ {+1} $. These definitions guarantee phase segregation.

\begin{figure}[!b]
\centering
  \includegraphics[width=0.9\textwidth]{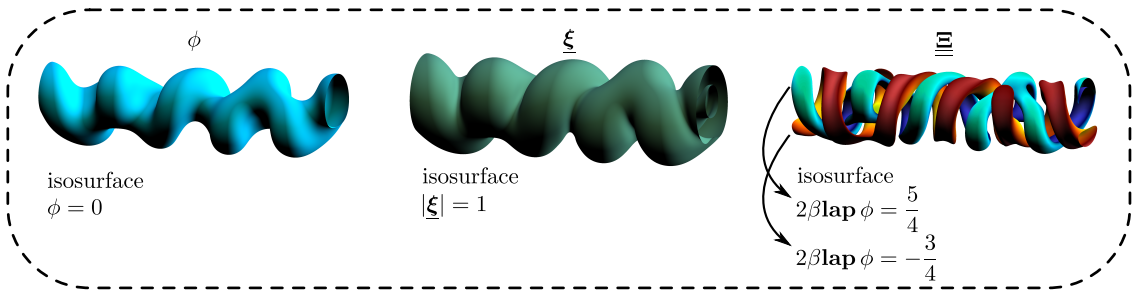}
  \caption{(Color online) From left to right: order-parameter, first-order microstress, and second-order microstress (colored by the order-parameter) at $ {t = 10} $.}
\label{fg:1}
\end{figure}

Figure \ref{fg:1} depicts the isosurfaces of $ {\phi} $, $ {\T{\xi}{1}} $, and $ {\T{\Xi}{2}} $ at $ {t = 10} $. Here, one can observe the different structures of $ {\T{\xi}{1}} $ and $ {\T{\Xi}{2}} $ at early stages, represented by isosurfaces. While $ {\T{\xi}{1}} $ captures the effects along a lattice near the interface, $ {\T{\Xi}{2}} $ accounts for nonlocal effects in the neighborhood of the lattice. Roughly speaking, $ {-\divx{\T{\Xi}{2}}} $ provides, in an averaged sense, the interaction among neighboring lattices.

\begin{figure}[!]
  \centering
  \subfloat[$ {t = 1} $]{\label{fg:t1}\includegraphics[width=0.25\textwidth]{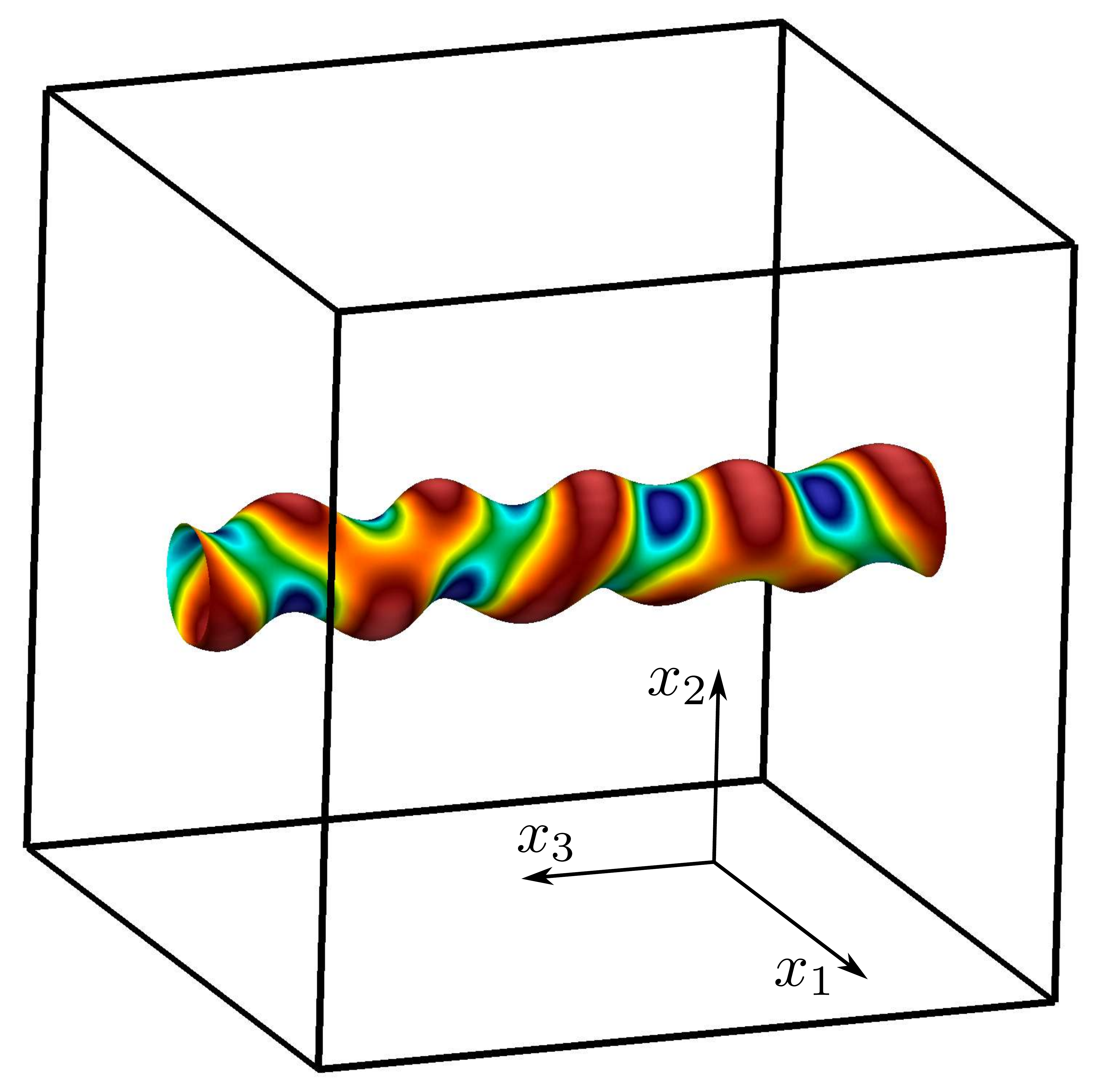}}
  \subfloat[$ {t = 10} $]{\label{fg:t10}\includegraphics[width=0.25\textwidth]{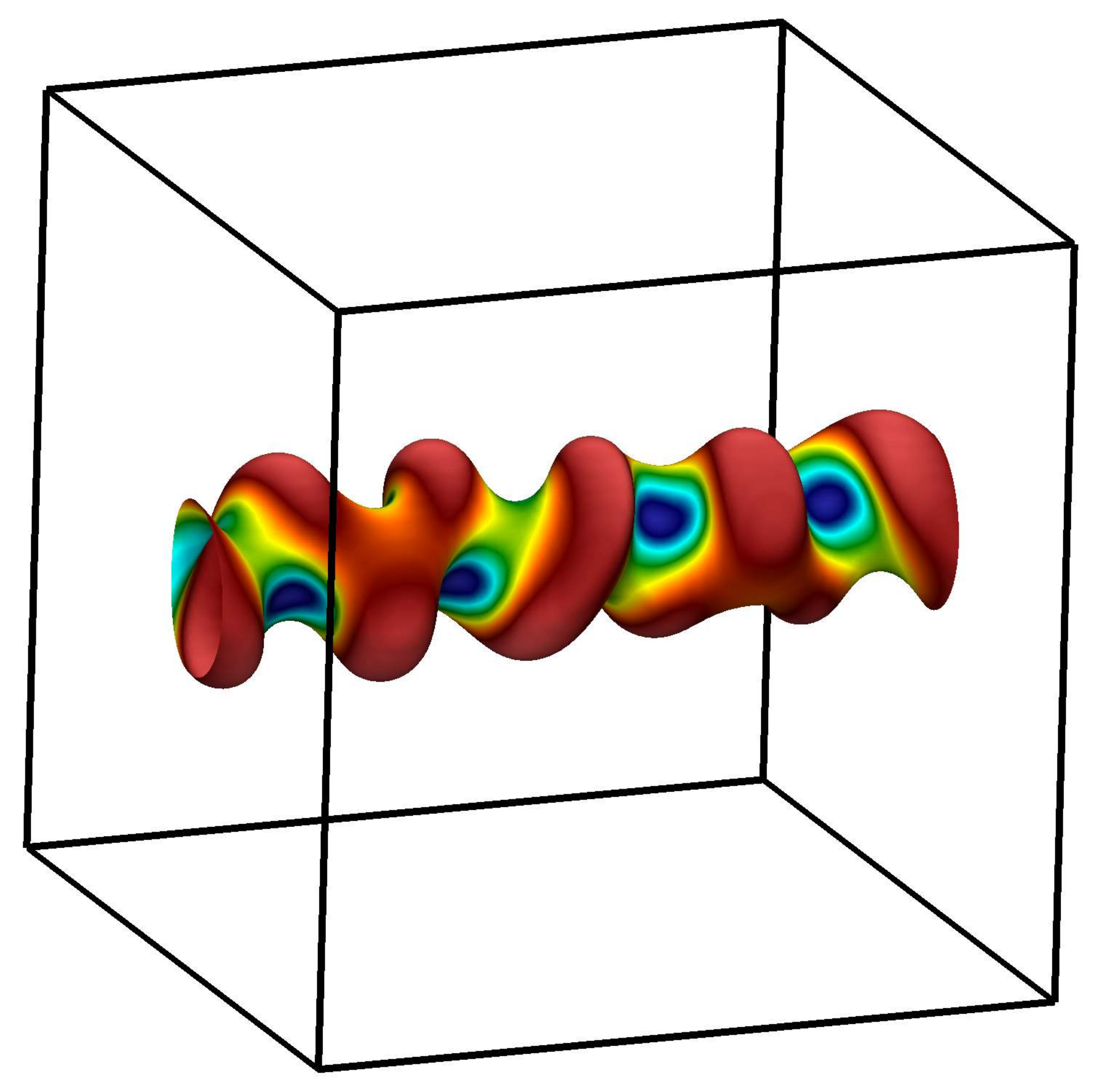}}
  \subfloat[$ {t = 20} $]{\label{fg:t20}\includegraphics[width=0.25\textwidth]{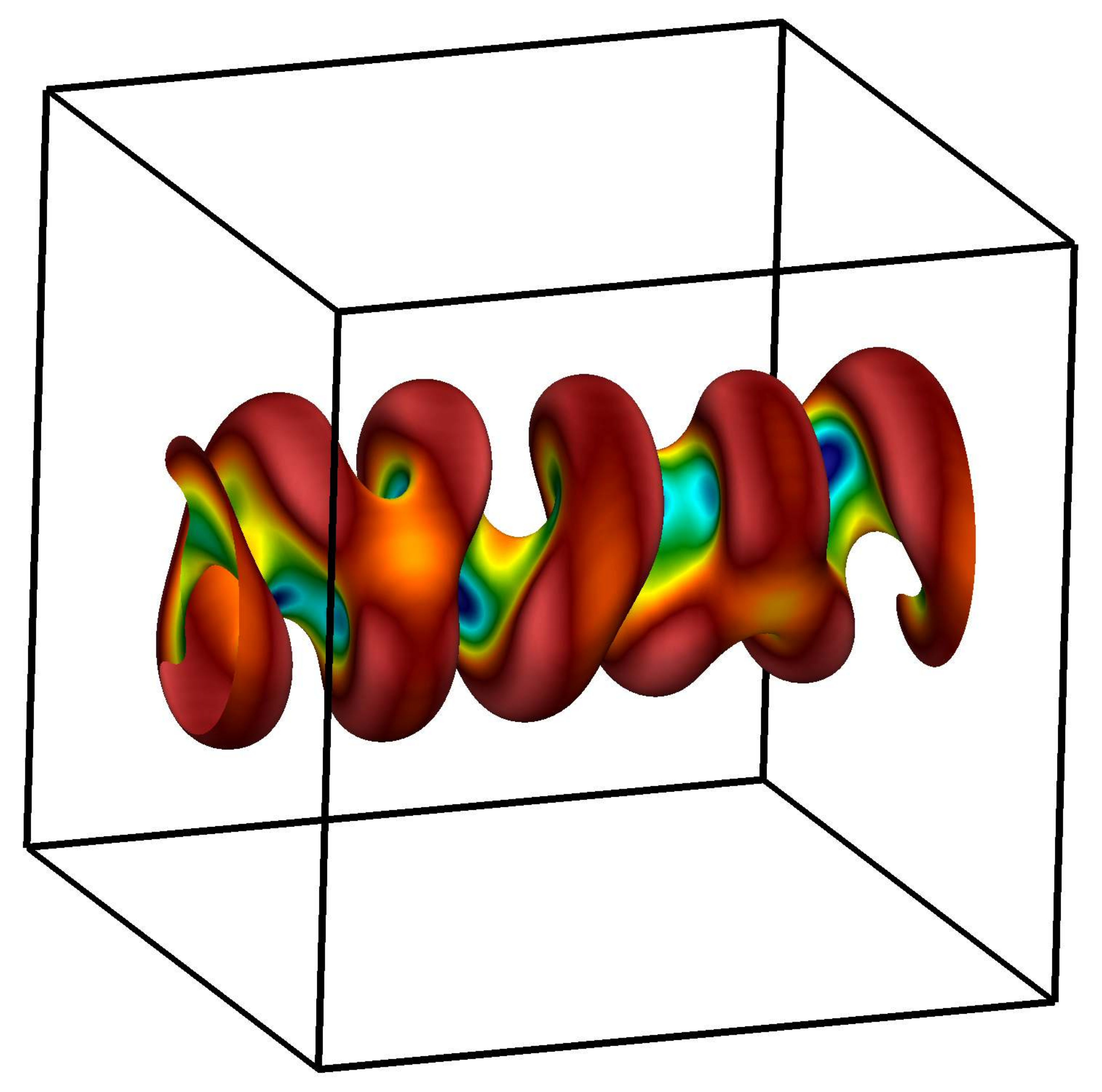}} \\
  \subfloat[$ {t = 30} $]{\label{fg:t30}\includegraphics[width=0.25\textwidth]{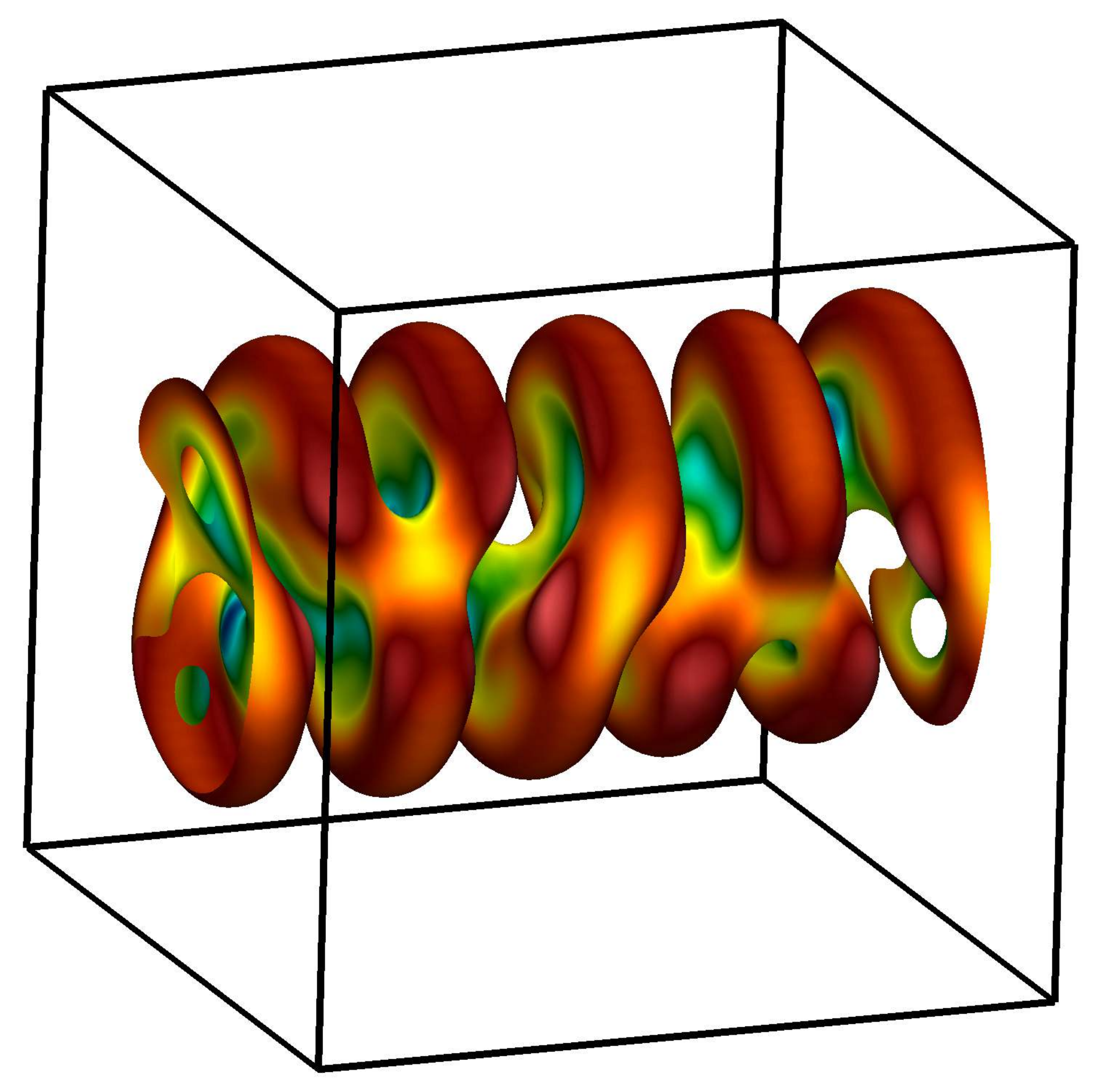}}
  \subfloat[$ {t = 50} $]{\label{fg:t50}\includegraphics[width=0.25\textwidth]{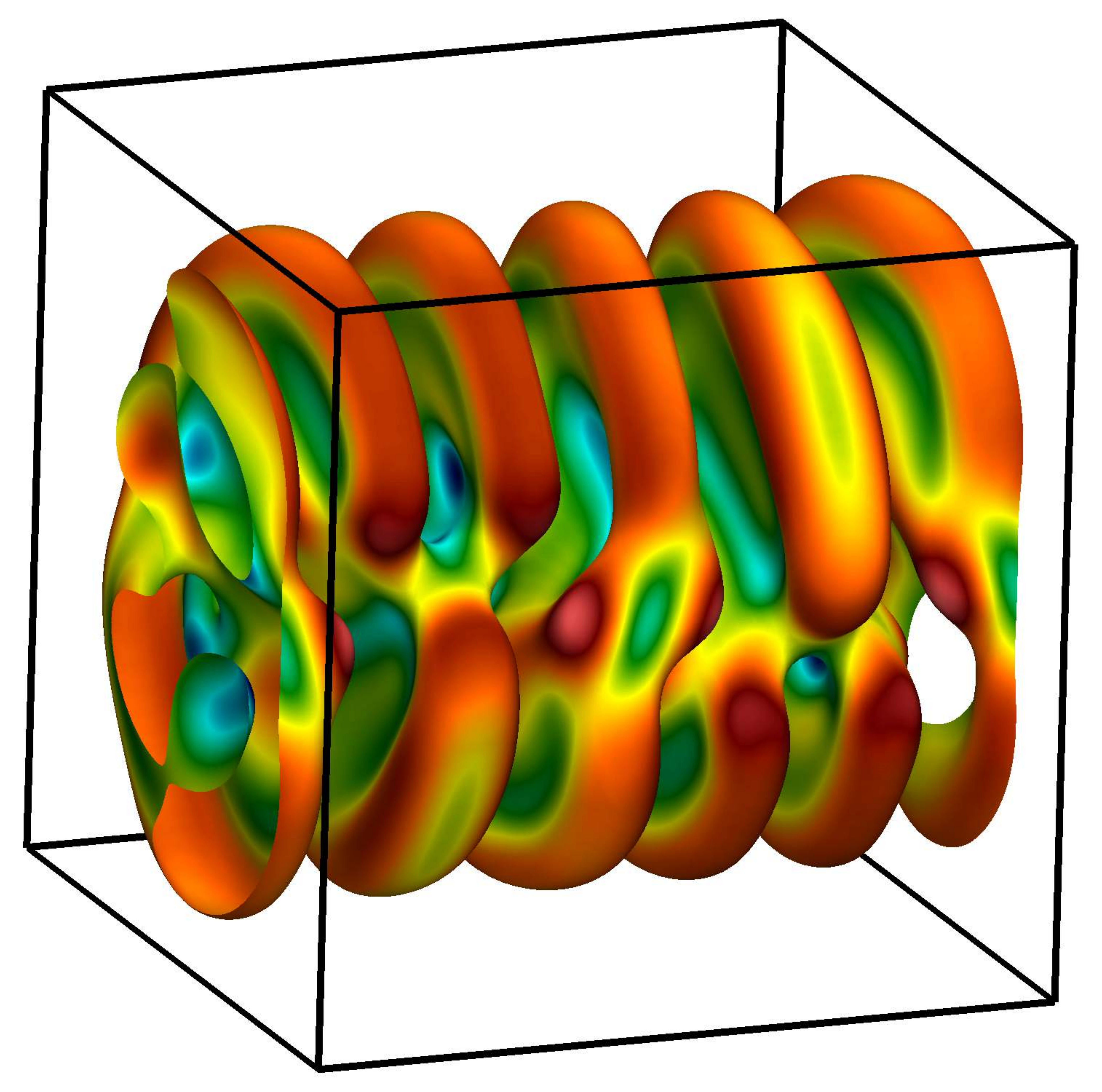}}
  \subfloat[$ {t = 145} $]{\label{fg:t145}\includegraphics[width=0.25\textwidth]{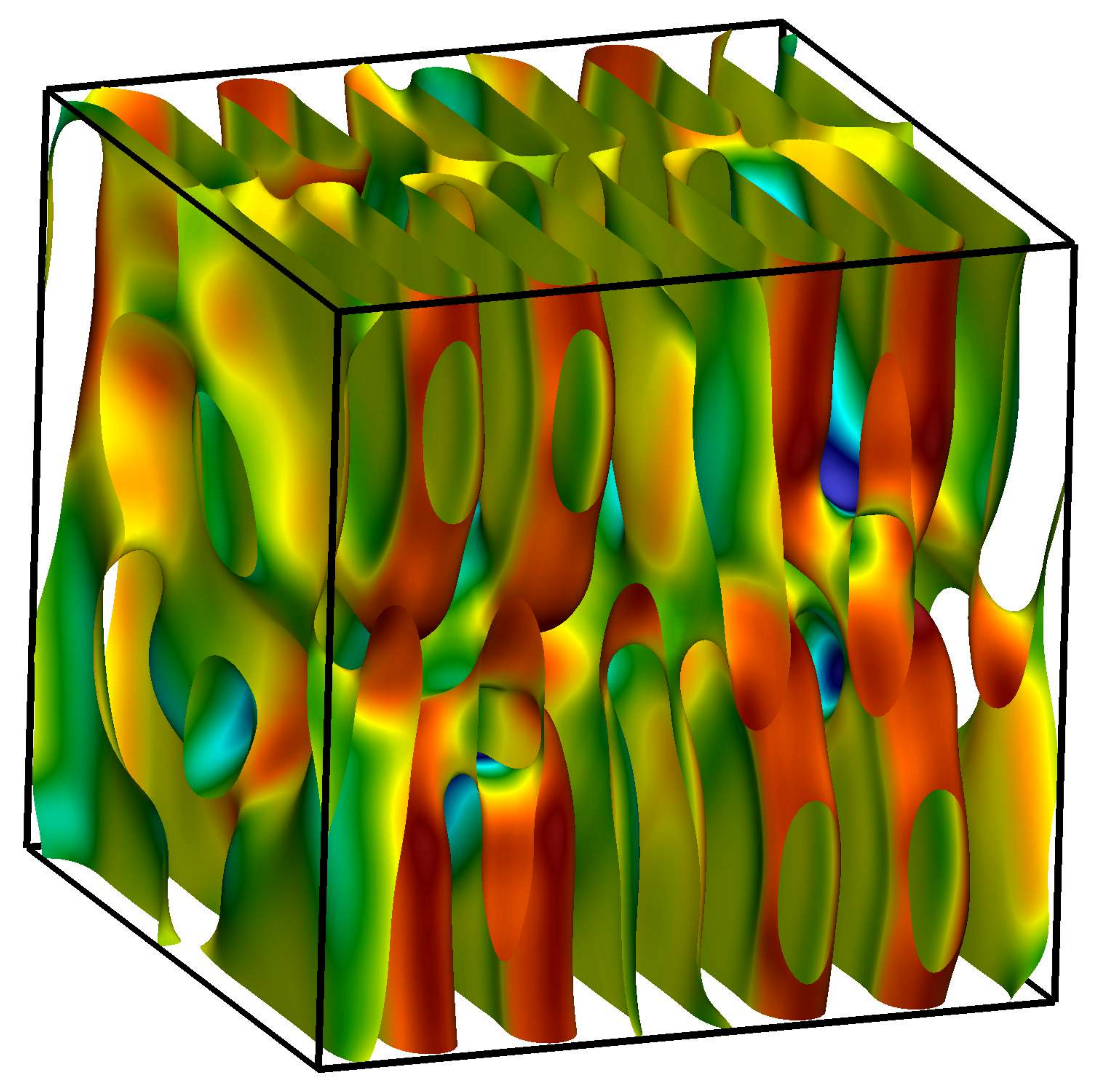}}
  \caption[(Color online) Evolution of the order-parameter $ {\phi} $ colored by its Laplacian]{(Color online) Evolution of the order-parameter $ {\phi} $ colored by its Laplacian.}
  \label{fg:2}
\end{figure}

Figure \ref{fg:2} shows the evolution of $ {\phi} $ colored by its Laplacian from $ {t = 1} $ to $ {t = 145} $. Here, we identify some symmetries in the structure of the phase-field that are preserved in time. For instance, at $ {x_3 = 10} $ and $ {x_3 = 30} $, both $ {\T{\xi}{1}} $ and $ {\divx{\T{\Xi}{2}}} $ do not have an out-of-plane component, meaning that the wavelength in $ {x_3} $ direction is $ {20} $. There is also a $ {\pi \, rad} $ rotational symmetry around $ {x_1} $. Additionally, the simulation is free of numerical oscillations (as shown by \cite{VIG16b} numerical oscilations can yield nonphysical solutions) and no numerical dissipation is used.
\begin{figure}[!]
\centering
  \includegraphics[width=0.8\textwidth]{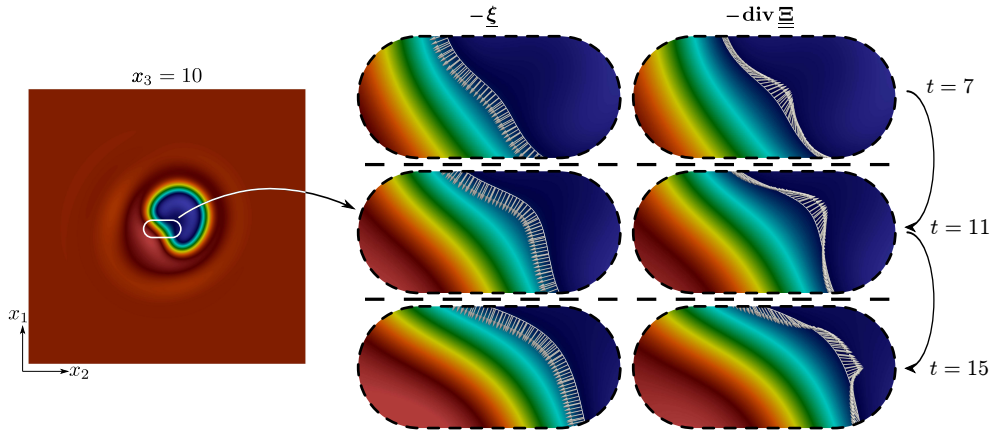}
  \caption{(Color online) Detail of the microstresses evolution in the plane $ {x_1 = 10} $}
\label{fg:3}
\end{figure}

Figure \ref{fg:3} depicts a slice at $ {x_3 = 10} $ at early stages to show first- and second-order microstresses over the isovalue surface of $ {\phi = -1.2} $. As shown by \cite{ESP16}, $ {\T{\xi}{1}} $ and $ {-\divx{\T{\Xi}{2}}} $ have a positive inner product on their overall behavior if the coefficients $ {\gamma} $ and $ {\beta} $ have the same sign, and they are parallel if the isosurfaces of $ {\phi} $ have a constant curvature. In the detail of this figure, we show the microstresses evolution from $ {t = 7} $ to $ {t = 15} $; there is a deviation of $ {-\divx{\T{\Xi}{2}}} $ from $ {-\T{\xi}{1}} $, where the larger the curvature changes the larger the deviation from one another.

\begin{figure}[!t]
\centering
  \subfloat[Plane $ {x_1 x_3} $]{\label{fg:pl1}\includegraphics[width=0.25\textwidth]{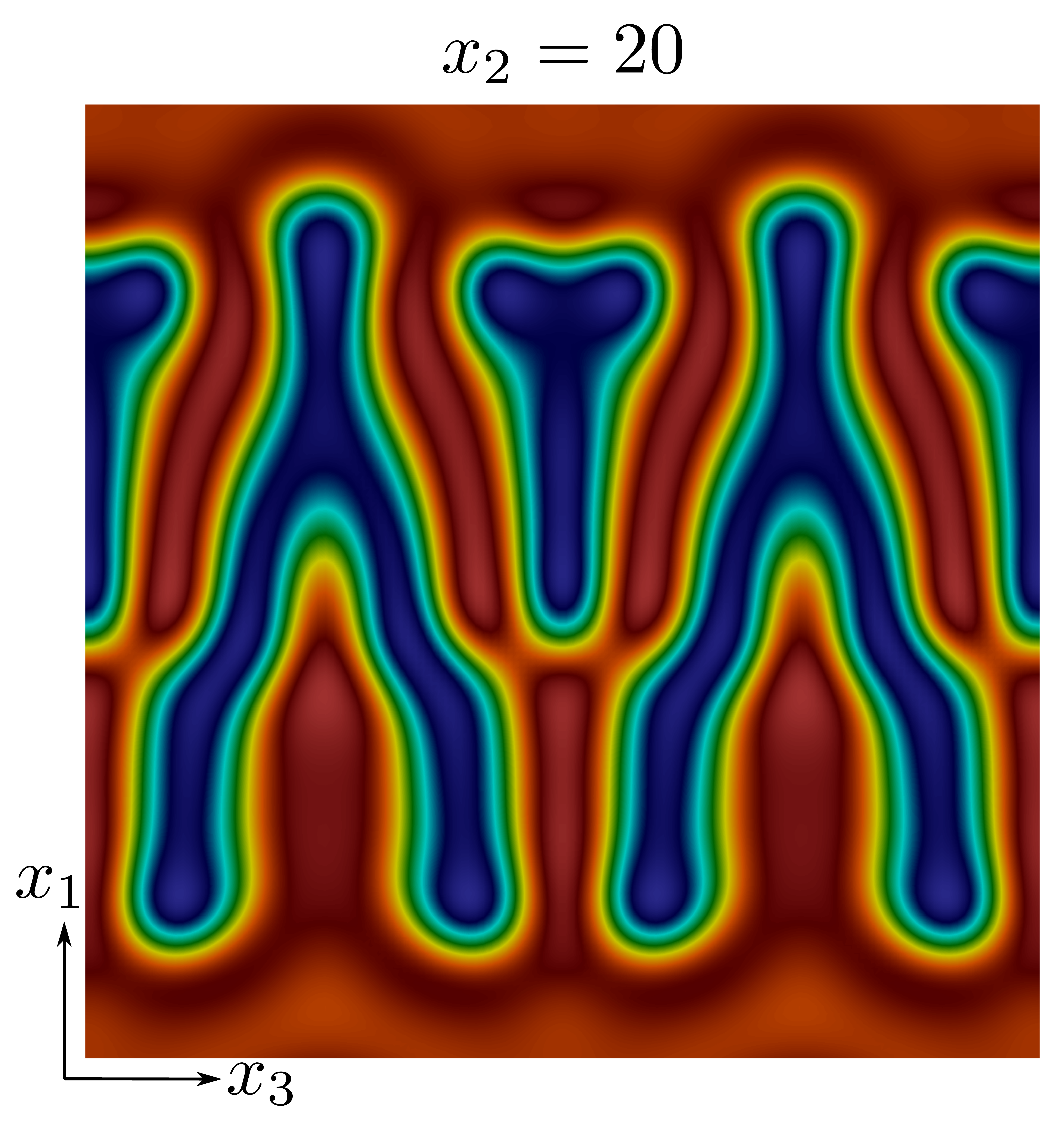}}
  \subfloat[Plane $ {x_1 x_2} $]{\label{fg:pl2}\includegraphics[width=0.25\textwidth]{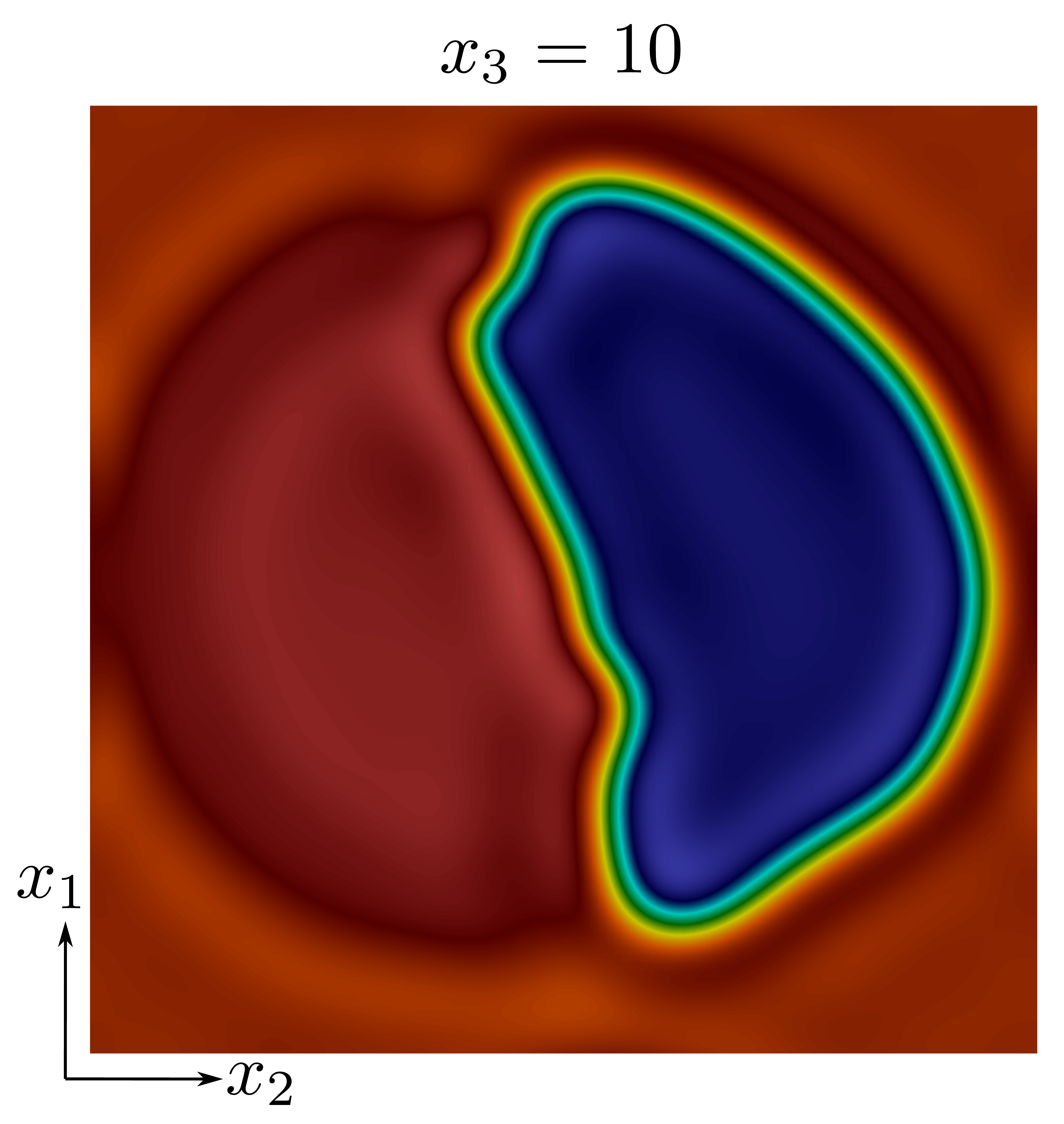}}
  \caption[(Color online) On the left a slice $ {x_2 = 20} $. On the right a slice $ {x_3 = 10} $ at $ {t = 50} $]{(Color online) On the left a slice $ {x_2 = 20} $. On the right a slice $ {x_3 = 10} $ at $ {t = 50} $.}
\label{fg:4}
\end{figure}

Figure \ref{fg:4} shows two slices of the domain. On the left, a slice at $ {x_2 = 20} $ depicts several smaller ordered structures, whereas on the right, a slice at $ {x_3 = 10} $ depicts two large structures at $ {t = 50} $.

%===============================================================================================================%

\section{Conclusions}\label{sec:conclusions}

We analyze the thermodynamics of the Swift--Hohenberg theory. Our derivation is based on a microbalance between the microforce and the first- and second-order microstresses. In the Swift--Hohenberg theory, we obtain an effective microstress $ {\T{\xi}{1} - \divx{\T{\Xi}{2}}} $ depending on a first- ($ {\T{\xi}{1}} $) and second-order ($ {\T{\Xi}{2}} $) microstresses. After explicitly stating the first and second laws of thermodynamics for this model, we generalize the model and detail some simple parameter choices. Finally, a highly-resolved simulation shows the interplay between the first- and second-order microstresses.

%===============================================================================================================%

\section{Acknowledgments}\label{sec:acknowledgments}

This publication was made possible in part by the CSIRO Professorial Chair in Computational Geoscience of Curtin University, the National Priorities Research Program grant \mbox{7-1482-1-278} from the Qatar National Research Fund (a member of the Qatar Foundation), and by the European Union's Horizon 2020 Research and Innovation Program of the Marie Sk\l{}odowska-Curie grant agreement No. 644202, the J. Tinsley Oden Faculty Fellowship Research Program at the Institute for Computational Engineering and Sciences (ICES) of the University of Texas at Austin has partially supported the visits of VMC to ICES, the Spring 2016 Trimester on ``Numerical methods for PDEs'', organized with the collaboration of the Centre Emile Borel at the Institut Henri Poincare in Paris supported VMC's visit to IHP in October, 2016.

%===============================================================================================================%

\bibliographystyle{elsarticle-harv}
%\bibliography{bib}

\end{document}